\documentclass[aps,pre,twocolumn,showpacs,superscriptaddress,groupedaddress]{revtex4}
\pdfoutput=1
\usepackage{hyperref}
\usepackage{epsfig}
\usepackage{amsmath}
\usepackage{graphicx}       
\usepackage{wrapfig}
\usepackage{dcolumn}  
\usepackage{bm}     
\usepackage{amssymb} 
\usepackage{titlesec}
\usepackage{mathtools}
\usepackage{relsize}
\usepackage{cleveref}
\usepackage{comment}
\usepackage{epsf}
\begin{document}
\title{Epidemic fronts in complex networks with metapopulation structure}
\author{Jason Hindes}
\affiliation{Laboratory of Atomic and Solid State Physics, Cornell University, Ithaca, New York}
\author{Sarabjeet Singh}
\affiliation{Theoretical and Applied Mechanics, Sibley School of Mechanical and Aerospace Engineering, Cornell University, Ithaca, New York}
\author{Christopher R. Myers}
\affiliation{Laboratory of Atomic and Solid State Physics, Cornell University, Ithaca, New York}
\affiliation{Institute of Biotechnology, Cornell University, Ithaca, New York}
\author{David J. Schneider}
\affiliation{Plant-Microbe Interactions Research Unit, Robert W. Holley Center for Agriculture and Health, Agricultural Research Service, United States Department of Agriculture, Ithaca, New York}
\affiliation{ Department of Plant Pathology and Plant-Microbe Biology, Cornell University, Ithaca, New York}
\begin{abstract}

Infection dynamics have been studied extensively on complex networks, yielding insight into the effects of heterogeneity in contact patterns on disease spread. Somewhat separately, metapopulations have provided a paradigm for modeling systems with spatially extended and ``patchy" organization. In this paper we expand on the use of multitype networks for combining these paradigms, such that simple contagion models can include complexity in the agent interactions and multiscale structure. We first present a generalization of the Volz-Miller mean-field approximation for Susceptible-Infected-Recovered (SIR) dynamics on multitype networks. We then use this technique to study the special case of epidemic fronts propagating on a one-dimensional lattice of interconnected networks -- representing a simple chain of coupled population centers -- as a necessary first step in understanding how macro-scale disease spread depends on micro-scale topology. Using the formalism of front propagation into unstable states, we derive the effective transport coefficients of the linear spreading: asymptotic speed, characteristic wavelength, and diffusion coefficient for the leading edge of the pulled fronts, and analyze their dependence on the underlying graph structure. We also derive the epidemic threshold for the system and study the front profile for various network configurations. To our knowledge, this is the first such application of front propagation concepts to random network models. 

\end{abstract}
\maketitle

\section{\label{sec:Intro} INTRODUCTION}
Network theory has proven a powerful framework for studying the effects of randomness and heterogeneity on the dynamics of interacting agents with non-trivial connectivity patterns \cite{Strog1}. One of the most important applications of this work is to the spread of infectious diseases among human populations, where the interaction structure is highly complex,  showing salient features such as power-law degree distributions, small average path lengths, and modularity  \cite{Newman3,Vespignani1}. Various models have been proposed, primarily with random graph configuration, that incorporate these complex features while remaining theoretically tractable. Within the context of disease dynamics, graph nodes are generally taken to represent individuals, and edges to represent interactions between them, through which infection can spread. Both deterministic and stochastic infection dynamics have been studied on networks as well as bond percolation for the associated branching process  \cite{Newman2, Robins1,Noel1,Noel2,Volz1}. How various thermodynamic quantities of interest -- such as the steady state incidence, the epidemic (percolation) threshold, and the distribution of small outbreak sizes -- depend upon network topology is of great interest. 

Often these approaches disregard the multiscale organization of many real systems, in which agents can be most naturally thought of as partitioned into densely connected communities with sparser coupling among neighboring communities. In some cases, it may be useful to conceptualize the topology as a network of networks, where agent-to-agent interactions and community-to-community interactions are both useful representations depending on the scale of resolution \cite{Vazquez1}. The latter has been successfully developed in ecology, with a network of interconnected populations referred to as a ``metapopulation" \cite{Keeling1,Hanski1}. This framework is very useful in studying large scale propagation of diseases where most infection transmission occurs in localized regions, but can be transported on larger scales by the mobility of individuals, traveling among population centers \cite{Belik1}. However, most metapopulation models assume that populations are fully mixed, with no inherent complexity in the connectivity between agents. Much less understood is how the multiscale structure of agent interactions affects the larger scale propagation of infectious processes through interconnected networks \cite{Dikison,Mendiola}. 

In this paper, we expand on a possible avenue for addressing this question using a multitype generalization of random graphs with simple, meta-level topology \cite{Vazquez1, Antoine1}, and construct a dynamical mean-field theory for the SIR infection model in multitype configuration model networks. Putting these together, we analyze the average infection dynamics and propagating front profile on a simple metapopulation composed of coupled population centers on a one-dimensional lattice and calculate the phenomenological transport properties of the system as functions of the underlying network's degree distributions. Our results are compared to stochastic simulations of the infection kinetics on various networks and found to be in good agreement in the thermodynamic limit. Broadly, we present this work as an illustration of how well-developed ideas from different areas of statistical physics and ecology can be naturally combined.   

\section{\label{sec:MultiTypeNet} MULTITYPE  CONFIGURATION MODEL NETWORKS}
In order to incorporate relevant node attribute information into our network models, (generically applicable for such things as age, sex, ethnicity, and place of residence), we use a generalization of configuration model random graphs, wherein nodes are assigned a type from an arbitrary set of $M$ possible types and a degree to each type from an arbitrary joint distribution for degree types, $P_{i}(k_{1},k_{2},...k_{M})=P_{i}(\vec{k})$, with degree $k_j$ denoting the number of connections to nodes of type $j$ \cite{Strog1,Antoine1,Vazquez1}. Additionally, nodes of type $i$ occupy a fraction of the total network $w_{i}$, where $\sum_{i}w_{i} =1$. Following the configuration model prescription, we consider graphs chosen uniformly at random from the ensemble of possible graphs with the prescribed degree distributions and self-consistent edge constraint: $w_{i}\sum_{\vec{k}}k_{j}P_{i}({\vec k}) = w_{j}\sum_{{\vec{k'}}}k'_{i}P_{j}({\vec{k'}}), \forall (i, j)$\cite{Strog1,Newman1,Antoine1}.

From this formalism, a variety of quantities can be described compactly using generating functions \cite{Strog1,Wilf1}. The generating function for the probability of a randomly selected node of type $i$ to have degree ${\vec{k}}$, is given by
\begin{equation}
\label{eq:GF}
G_{i}({\vec{x}}) = \sum_{{\vec{k}}}P_{i}({\vec{k}})\prod_{l=1}^{M}x_{l}^{k_{l}} :   
\end{equation}
\noindent written as a power series in $\vec{x}$, an auxiliary variable defined over the unit interval, with expansion coefficients equal to the respective probabilities. Moments of the degree distributions can be represented simply as derivatives of the corresponding generating function. For example, the average degree of a type $i$ node to a type $j$ node is 
\begin{equation}
\label{eq:Avg}
\sum_{{\vec{k}}}k_{j}P_{i}({\vec{k}}) = \partial_{x_{j}}G_{i}({\vec{x}})\vert_{\vec{1}} \equiv \left<k_{j}\right>_{i}. 
\end{equation}
Since node interactions occur along edges, an important quantity in network models is the excess degree: the number of neighbors a node has which can be reached by selecting a randomly chosen edge, and not including the neighbor on the end of the selected edge. For a multiype configuration model network, the probability that a randomly chosen edge from a type $i$ node leads to a type $j$ node with degree $\vec{k}$ is proportional to $k_{i}P_{j}({\vec{k}})$, and thus the probability for the corresponding excess degree is generated by $\partial_{x_{i}}G_{j}({\vec{x}})/\partial_{x_{i}}G_{j}({\vec{x}})\vert_{\vec{1}}$  \cite{Antoine1}, with average degree to type $l$ nodes,  
\begin{equation}
\label{eq:Exc}
\left<k_{l}\right>_{i-j} = \frac{\partial_{x_{l}}\partial_{x_{i}}G_{j}({\vec{x}})\vert_{\vec{1}}}{\partial_{x_{i}}G_{j}({\vec{x}})\vert_{\vec{1}}}= \frac{\left<k_{l}k_{i}\right>_{j}}{\left<k_{i}\right>_{j}}-\delta_{il}. 
\end{equation}
By construction, this framework lacks two-point correlations, in which the excess degree distributions depend on the degrees of  both nodes sharing an edge \cite{Vespignani1}.

\section{\label{sec:VolzMiller} VOLZ-MILLER MEAN-FIELD SIR IN MULTITYPE NETWORKS}
In this report we consider simple dynamics for disease spread: the Susceptible-Infected-Recovered (SIR) model, wherein each individual is assigned a disease state, $Y\in \{S,I,R\}$, and may undergo reactions to other states depending on its state and the state of its neighbors. In this model, if a node of type $i$ is susceptible and has a single infected neighbor of type $j$, then it will change its state to infected with a constant probability per unit time $\beta_{ji}$. Likewise, an infected node of type $i$ will recover with a constant probability per unit time $\gamma_{i}$. Since the underlying dynamics is a continuous time Markov process, a complete analysis would describe the full probability distribution for all system trajectories. However for our purposes, it will be sufficient to focus on the behavior of extensive outbreaks (i.e., those which scale with the system size), the average dynamics of which, can be derived in the limit  when the number of nodes tends to infinity, by generalizing a mean-field technique for single type networks, developed by Volz and Miller, to multitype networks. Below, we follow the basic structure of the derivations presented in \cite{Volz2, Miller1}.

In the thermodynamic limit, configuration model random graphs are locally tree-like \cite{Newman1}, which by construction allows them to satisfy many of the generic criteria for the applicability of mean-field theory assumptions \cite{Gleeson1}. In our case, we assume that nodes are differentiated by their degree and disease state alone and that susceptible nodes feel a uniform force of infection along every edge, related to the average number of edges connecting susceptible and infected nodes at any given time in the network: a Curie-Weiss type approximation \cite{Goldenfeld}. Furthermore, from the perspective of susceptible nodes, all infection attempts along different edges can be treated as uncorrelated -- a consequence of the local tree-like property \cite{Newman1,Miller1, Robins1}-- and thus we assume that the states of neighbors of susceptible nodes are effectively independent.

Let the probability that a node of type $j$ has not transmitted the infection to a node of type $i$ along a randomly chosen $i-j$ edge, be $\theta_{ij}$. This quantity is interpretable as the complement of the average cumulative hazard function along such edges. Given $\theta_{ij}$, it follows that the fraction of susceptible nodes of type $i$ at  time $t$ is 
\begin{align}
\label{eq:SuscI}
S_{i}(t)&= \sum_{{\vec{k}}}P_{i}({\vec{k}})\prod_{j=1}^{M}\theta_{ij}^{k_{j}}(t)  = G_{i}(\theta_{i1}(t),\theta_{i2}(t),...\theta_{iM}(t)) \nonumber \\
            &\equiv G_{i}(\vec \theta_{i}(t)). 
\end{align}
\noindent The fractions of infected and recovered nodes of type $i$ follow from probability conservation, $S_{i}+I_{i}+R_{i}=1$, and a constant recovery rate for infected nodes $\gamma_{i}$: 
 \begin{align}
 \label{eq:IRequ}
 \frac{dI_{i}}{dt}&= - \frac{d\vec{\theta_{i}}}{dt} \cdot \vec{\nabla} G_{i}(\vec{x})\vert_{\vec \theta_{i}} - \gamma_{i}I_{i} \nonumber \\
 \frac{dR_{i}}{dt}&= \gamma_{i}I_{i} \;, 
\end{align}  
\noindent with the total fraction of susceptible nodes  
\begin{equation}
\label{eq:Susc}
 S=\sum_{i}w_{i}G_{i}(\vec \theta_{i})\equiv \vec{w} \cdot \vec{G}(\underline{\underline{\theta}}). 
\end{equation}

The central probability and order parameter, $\theta_{ij}$, can be subdivided into three compartments depending on the disease state of the terminal node $j$,  
\begin{equation}
\label{eq:ThetaFlux}
\theta_{ij} = \theta_{ij}^{S} + \theta_{ij}^{I} + \theta_{ij}^{R}\;, 
\end{equation}
\noindent and its dynamics determined by tracking the fluxes among these compartments. Since $\theta$ can only change when an infected node transmits the disease, the rate at which $\theta_{ij}$ changes is equal to the rate at which a corresponding neighbor infects, and therefore $d\theta_{ij}=-\beta_{ji}\theta_{ij}^{I}dt$. Similarly, since $\theta^{R}$ can only change if an infected node recovers, the rate at which  $\theta_{ij}^{R}$ changes is equal to the rate at which a corresponding neighbor recovers, and thus $d\theta_{ij}^{R}=\gamma_{j}\theta_{ij}^{I}dt$. Lastly, the probability that a type $j$ neighbor of a type $i$ node has not transmitted and is susceptible, $\theta_{ij}^{S}$, is simply the probability that the corresponding neighbor is susceptible. Because this neighbor could not have been infected along any of its other edges and has excess degree distribution generated by  $\partial_{x_{i}}G_{j}({\vec{x}})/\partial_{x_{i}}G_{j}({\vec{x}})\vert_{\vec{1}}$ , it follows that  $\theta_{ij}^{S}=\partial_{x_{i}}G_{j}({\vec{x}})\vert_{\vec{\theta_{j}}}/\partial_{x_{i}}G_{j}({\vec{x}})\vert_{\vec{1}}$. Combining the latter with the two flux relations and the initial conditions  \eqref{eq:ThetaFlux}, $\theta_{ij}(0)=1$ and $\theta_{ij}^{R}(0)=0$, we find 

\begin{align}
\label{eq:MFequ}
&\frac{d\theta_{ij}}{dt} = \beta_{ji}\left( \frac{\partial_{x_{i}}G_{j}(\vec x)\vert_{\vec \theta_{j}}}{\partial_{x_{i}}G_{j}(\vec x)\vert_{\vec{1}}}  -\theta_{ij}\right) + \gamma_{j}\left(1-\theta_{ij}\right).   \nonumber \\
\end{align}
\noindent These $M^{2}$, first-order and coupled ODEs, $\dot{\underline{\underline{\theta}}}=\underline{\underline{F}}(\underline{\underline{\theta}})$, define the full system's approximate mean dynamics, and form the basis of our subsequent analysis. For a more detailed derivation of the analogous results for the special case of a single type network, see \cite{Volz2, Miller1}. 

The steady state is given by the fixed point of \eqref{eq:MFequ},
\begin{align}
\label{eq:SteadyState}
&\bar \theta_{ij} = (1-T_{ji}) +T_{ji}\frac{\partial_{x_{i}}G_{j}(\vec{x})\vert_{\vec{\bar{\theta_{j}}}}}{\partial_{x_{i}}G_{j}(\vec{x})\vert_{\vec{1}}} \; , \nonumber \\ 
\end{align}
\noindent which upon substitution into \eqref{eq:SuscI}, gives the cumulative infection,  $P=1-S$, at equilibrium (i.e., the final epidemic size), with $T_{ji}= \beta_{ji}/\left(\beta_{ji}+\gamma_{j}\right)$ the corresponding bond percolation probability, or transmissibility \cite{Antoine1}. This can have a non-trivial solution corresponding to the existence of extensive outbreaks, if the disease-free state, $\theta_{ij}=1 \;\; \forall (i, j)$, is unstable. The threshold or phase transition, which signifies the region in parameter space that separates the epidemic and non-epidemic phases, can be obtained through a stability analysis of the disease-free state, where the eigenvalue of the Jacobian for \eqref{eq:MFequ} with the largest real part, is real and vanishes when  
\begin{align}
\label{eq:GenThresh}
&det(\underline{\underline{N}}-\underline{\underline{I}})=0, \;\;\; \text{with} \\ \nonumber 
&N_{(i,j)(k,l)}= T_{ji}\delta_{jk}\left<k_{l}\right>_{i-j}  
\end{align}
\noindent an $M^{2}$x$M^{2}$ matrix \cite{Meyer}. Similar results for the equilibrium properties are derivable from a multitype bond percolation approach \cite{Antoine1}.

\section{\label{sec:MultiScaleFramework}FRAMEWORK FOR MULTISCALE NETWORKS}
\begin{figure}[htbp]
\begin{center}
\includegraphics[scale=0.34]{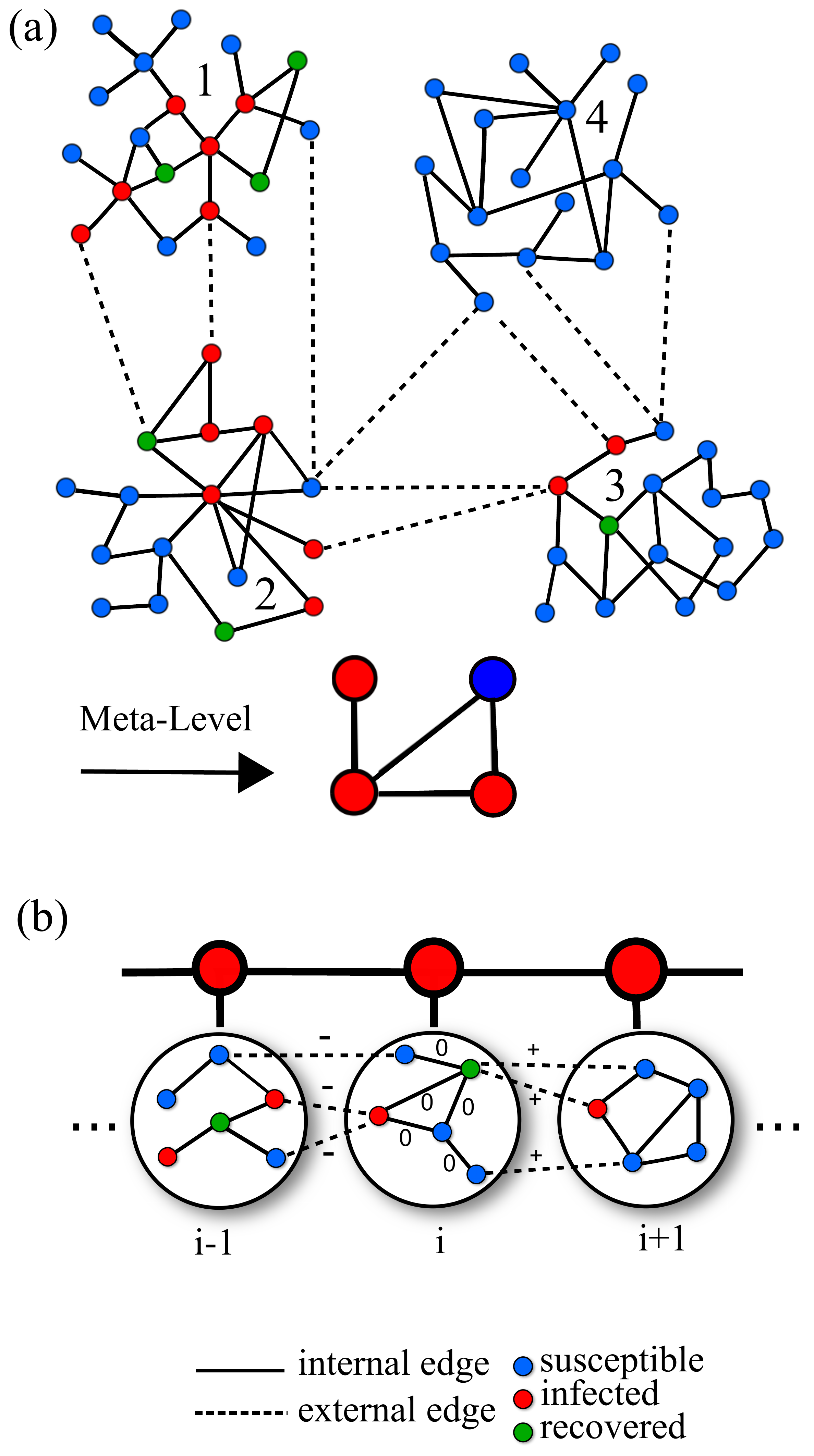}
\caption{ (a) A schematic of SIR dynamics on a metapopulation, where infection spreads along edges connecting nodes of various types at the finest scale (shown with integer, population labels), and the macro-scale topology identifies which populations are directly connected through agent interactions. (b) A particular example of this framework, in which the macro-scale topology takes the form of a one-dimensional lattice. In Sec.\ref{sec:1DLatt}, we focus on a simple case with configuration model construction, where each site has an identical degree distribution, specifying the probability of having a given number of internal (0), right (+), and left (-) external edges (shown above with labels for site $i$).}
\label{fig:Schematic}
\end{center}
\end{figure}

Of interest to us are systems where type structure adds an additional scale of relevant topology, and not just demographic complexity \cite{Vazquez1, Antoine1}. For instance, we can apply the multitype network formalism to a simple model for a metapopulation by affiliating population centers with node types and coupling among populations with edges connecting their constituent nodes. In this way, a complex topology can be encoded on a micro-scale with a macro-scale adjacency matrix, $\underline{\underline{A}}$, describing which populations are directly connected through node interactions \cite{Vazquez1}. We envisage example systems where $\underline{\underline{A}}$ describes the connectivity among urban centers, such as cities, towns, or villages, facilitated by roads or airlines. By conceptualizing the topology in this manner, we can study the phenomenology of infection propagation among population centers and describe how the propagation properties depend on the underlying connectivity patterns. A schematic is shown in Fig.\ref{fig:Schematic}-(a) for a simple system with the pertinent structure. More broadly, we advance this approach as an avenue for combining the frameworks of network theory, metapopulations, and front propagation, which will be particularly useful if the interaction topology is coherent after some level of coarse graining. 

\section{\label{sec:1DLatt} 1-D LATTICE METAPOPULATION DYNAMICS}

To illustrate this approach, we consider a special case of the above where the macro-scale topology is an infinite one-dimensional lattice,  $M\rightarrow \infty$, in which agents interact with other agents of the same type and agents of neighboring types, $A_{nj}=(\delta_{j,n}+\delta_{j,n+1}+\delta_{j,n-1})$ \cite{Belik1}. If infection is started at a single site (e.g., site $0$) in a fully susceptible system, a strict directionality applies:  in order for site $i$ to be infected, sites $i-1, i-2,...$ must be infected first. In such a case, we expect a well-defined infection front to propagate through the lattice. In keeping with the above, we focus on an effective force of infection model among populations with static configuration networks having prescribed degree distributions -- a generalization of the paradigmatic, spatial SIR model in one dimension, where the assumption of well mixed populations is relaxed to include complexity in agent interactions \cite{Sazonov1, Keeling1}. A schematic is shown in Fig.\ref{fig:Schematic}-(b). 

Since each node has three edge types, the mean equations of motion describe a three-component field, $\vec \theta_{n}(t)\equiv\left(\theta_{nn},\theta_{nn+1},\theta_{nn-1}\right)\equiv\left(\theta^{0}_{n}(t),\theta^{+}_{n}(t),\theta^{-}_{n}(t)\right)$, where $(0)$, $(+)$, and $(-)$ denote internal, right-external, and left-external edges, at the corresponding site. For simplicity, homogeneity is  assumed, with $\beta$, $w$, $\gamma$, and $G$ all uniform -- reducing the field equations to
\begin{align}
\label{eq:1DLatt}
&\frac{d\theta^{0}_{n}}{d\tau}= (1-\theta^{0}_{n}) +T\left( \frac{G_{0}(\vec \theta_{n})}{G_{0}(\vec1)}   -1\right) \nonumber \\
&\frac{d\theta^{\pm}_{n}}{d\tau}= (1-\theta^{\pm}_{n}) +T\left( \frac{G_{\mp}(\vec \theta_{n\pm1})}{G_{\mp}(\vec1)}   -1\right),
\end{align}
\noindent where the time, $\tau$, is measured in units of $1/(\beta +\gamma)$, and the subscript in $G$ denotes a partial derivative with respect to the corresponding variable. For edge number consistency, $G_{+}(\vec{1})=G_{-}(\vec{1})$, but in general we allow for other asymmetries in the degree distributions. 

 \subsection{\label{sec:Dispersion}  Dispersion Relation and Transport Coefficients}
To understand the spatio-temporal dynamics \eqref{eq:1DLatt}, we first quantify how perturbations away from the unstable state propagate by linearizing the dynamics around the disease-free equilibrium, $\vec \theta_{n}(t)=\vec{1}-\vec{\epsilon}_{n}(t)$, and decoupling the perturbations into basis modes using the Inverse Discrete Fourier Transform, $\vec{\epsilon}_{n}(t)= \frac{1}{M}\sum^{M-1}_{\nu=0} \vec{\epsilon}_{\nu}(t)e^{i(2\pi\nu n/M)}$, (IDFT). The dispersion relation can be found by substituting the  IDFT into \eqref{eq:1DLatt}  and using the basis properties of orthogonality and completeness. In the limit $M\rightarrow \infty$ the site perturbations approach the integral, $\vec{\epsilon}_{n}(t)= \frac{1}{2\pi}\int_{0}^{2\pi} \vec{\epsilon}(k)e^{i\left(kn-\omega(k)t\right)}dk$. 

With this prescription, we find the dispersion relation takes the form of a cubic, characteristic equation
\begin{align}
\label{eq:Dispersion}
&\det\left(\underline{\underline{K_{e}}}(q)-\frac{1\!+\!s(q)}{T}\underline{\underline{I}}\right) = 0, \;\;\;\;\;\; \text{with}  \\ 
&\underline{\underline{K_{e}}}(q)= 
\renewcommand{\arraystretch}{2.5}
\setlength\arraycolsep{7.2pt}
\begin{bmatrix*}[l]
\large
\frac{\left<k_{0}^2\right>}{\left<k_{0}\right>}-1 &\frac{\left<k_{0}k_{+}\right>}{\left<k_{0}\right>}&\frac{\left<k_{0}k_{-}\right>}{\left<k_{0}\right>} \\
\frac{\left<k_{-}k_{0}\right>}{\left<k_{-}\right>}e^{-q}&\frac{\left<k_{-}k_{+}\right>}{\left<k_{-}\right>}e^{-q} &\left(\frac{\left<k_{-}^2\right>}{\left<k_{-}\right>}\!-\!1\right)e^{-q}\\
\frac{\left<k_{+}k_{0}\right>}{\left<k_{+}\right>}e^{q} &\left(\frac{\left<k_{+}^2\right>}{\left<k_{+}\right>}\!-\!1\right)e^{q} &\frac{\left<k_{+}k_{-}\right>}{\left<k_{+}\right>}e^{q}\\
\end{bmatrix*}  \nonumber 
\end{align}
\noindent which for convenience, is written in terms of $s$ and $q$, where $\omega=is$ and $k =iq$. Interestingly, this method reveals a generalization of the average excess degree matrix, $\underline{\underline{K_{e}}}(0)$ -- whose elements are found by selecting a randomly chosen edge in a particular direction, and counting the average number of reachable neighbors of a particular type -- for the interconnected network system, $\underline{\underline{K_{e}}}(q)$, which incorporates the relative states of adjacent sites on the lattice for each mode $q$. We expect this operator to emerge in similar problems on interconnected networks. 

Combining the above with the behavior of infection near the phase transition, where there is no exponential growth in time and each site has the same field value: $s\rightarrow0$ and $q\rightarrow0$ \eqref{eq:Dispersion},  we find a simple condition for the critical transmissibility $T_{c}$:
\begin{align}
\label{eq:1DThresh}
&T_{c}=\frac{1}{\lambda_{m}^{k}(0)}\;,
\end{align}
\noindent where $\lambda_{m}^{k}(q)$ is the maximum eigenvalue of $\underline{\underline{K_{e}}}(q)$, with $\lambda_{m}^{k}(0)$ corresponding to $\underline{\underline{K_{e}}}(0)$. Because the addition of external edges increases the spreading capacity of the disease, the critical transmissibility in the coupled system is less than the uncoupled case, implying that transport-mediated infections from neighboring sites can sustain epidemics even when individual populations on their own cannot \cite{Mendiola, Dikison}.  
 
Also from the dispersion relation, we can find the asymptotic transport coefficients for rightward moving disturbances by making a standard saddle-point-approximation of the perturbations' integral representation in fourier space: expanding the integrand around its dominant contribution, $k^{*}$, in the co-moving frame, $\xi=n-v^{*}t$,
\begin{align}
&e^{\!^{\mathlarger{i(kn\!-\!\omega(k)t)}}}\!\!\sim\!e^{\!\mathlarger{^{\!ik\xi\!}}}e^{it\left(\!kv^{*}\!-\omega(\!k^{*}\!)-\!\left.\mathlarger{\frac{d\omega}{dk}}\right\vert_{k^{*}}\!\!\!\left(k-k^{*}\right)\!\right)}\!e^{\!\!\left.\mathlarger{\frac{-it\left(k-k^{*}\right)^{2}}{2}}\!\mathlarger{\frac{d^{2}\!\omega}{dk^{2}}}\right \vert_{k^{*}}}\!\!\!
\nonumber
\label{eq:FourierArg}
\end{align}
\noindent and taking the infinite time limit while enforcing approximate constancy with no exponential growth and $\xi$ finite -- where $v^{*}$ is the asymptotic speed at which perturbations to the unstable state propagate \cite{Saarloos1}. This procedure uncovers an exponential moving pulse for the leading-edge of the infection profile with a diffusive correction \cite{Saarloos1}:
\begin{equation}
1-\theta  \sim \dfrac{e^{-q^{*}\xi}e^{-\xi^{2}/4D^{*}t}}{\sqrt{D^{*}t}}\;, 
\label{eq:PertField}
\end{equation}
\noindent where $q^*$, $v^*$, and $D^*$ satisfy the saddle-point relations: 
\begin{equation}
\label{eq:SaddlePoint}
v^{*}= \left. \frac{ds}{dq} \right \vert_{q^{*}} =\frac{s(q^{*})}{q^{*}} = T \left. \frac{d\lambda^{k}_{m}}{dq} \right \vert_{q^{*}}  = \frac{-1+T\lambda^{k}_{m}(q^{*})}{q^{*}} 
\end{equation}
\begin{equation}
\label{eq:Diff}
\text{and} \;\;\;\; D^{*}= \left. \frac{1}{2}\frac{d^{2}s}{dq^{2}} \right \vert_{q^{*}} = \left. \frac{T}{2}\frac{d^{2}\lambda^{k}_{m}}{dq^{2}} \right \vert_{q^{*}} \;,   
\end{equation}
\noindent giving a transcendental equation for $q^{*}$. When the average excess degree matrix is irreducible (the domain of interest to us), the dominant growth exponent for each $q$ is real and corresponds to a uniquely positive eigenvector \eqref{eq:Dispersion}  \cite{Meyer}, and thus we expect the same selected velocity for all fields \cite{Theodorakis}. Approximately, the fields propagate in this regime with proportions $\vec{1}-\epsilon(t)\vec{Q}(q^{*},s^{*})$, where $\vec{Q}(q^{*},s^{*})$ is the corresponding mode of $\underline{\underline{K_{e}}}(q^{*})$\eqref{eq:Dispersion}. If multiple solutions exist for $v^{*}$, the fastest solution is selected \cite{Saarloos1}. The characteristic wavelength, $1/q^{*}$, is related to the asymptotic size of the front's leading edge, and diverges near the phase transition. The diffusion coefficient, $D^{*}$, gives the effective widening of the mean-field pulse in the co-moving frame and is proportional to the largest finite-size correction to $v^{*}$ in the limit where the number of nodes at each site tends to infinity. 

In order to uncover the principal dependencies of the transport coefficients, we study \eqref{eq:Dispersion}-\eqref{eq:Diff} near the phase transition, where the power series expansion for the dispersion relation is a convenient representation; the latter is found by substituting $(s(q)+1)/T = a+bq+\frac{c}{2}q^{2}+...$ into \eqref{eq:Dispersion}, and equating powers in $q$. When $s^{*}$ and $q^{*}$ are small in the vicinity of $T_{c}$, the expansion can be truncated at low order, giving a Fisher-Kolmogorov-like dispersion relation with the approximate scaling 
\begin{equation*}
\begin{aligned}[c]
&s^{*}\sim\left(T\lambda^{k}_{m}(0)-1\right) \\
&q^{*}\sim\left(T\lambda^{k}_{m}(0)-1\right)^\frac{1}{2}{D^{*}}^{-\frac{1}{2}}\\
\end{aligned}
\qquad
\begin{aligned}[c]
&v^{*}\sim\left(T\lambda^{k}_{m}(0)-1\right)^\frac{1}{2}{D^{*}}^{\frac{1}{2}}\\
&\frac{D^{*}}{T\lambda^{k}_{m}(0)}\sim \delta
\end{aligned}
\end{equation*}
\begin{align}
\label{eq:Scaling}
&\delta=\frac{\frac{\left<k_{-}k_{0}\right>\left<k_{0}k_{+}\right>}{\left<k_{-}\right>\left<k_{0}\right>}+\frac{\left<k_{-}k_{+}\right>}{\left<k_{-}\right>} \left(\lambda^{k}_{m}(0)-\frac{\left<k_{0}^2\right>}{\left<k_{0}\right>}\!+\!1\right)}{\left(\lambda^{k}_{m}(0)-\lambda^{k}_{2}(0)\right) \left(\lambda^{k}_{m}(0)-\lambda^{k}_{3}(0)\right)}\;, 
\end{align} 

\noindent where $\lambda^{k}_{2}(0)$ and $\lambda^{k}_{3}(0)$ are the subdominant eigenvalues of $\underline{\underline{K_{e}}}(0)$. In this regime, we find an effective reaction-diffusion behavior with the generic dependence of the shape and speed of the propagating front's leading edge on the reproductive number $T\lambda^{k}_{m}(0)$ (a product of the spreading capacity along edges and the magnitude of topological fluctuations) and on the normalized diffusion coefficient $\delta$: measuring the relative strength of connection between lattice sites \eqref{eq:Scaling}. We see that the effective reaction rate is equal to the distance from the phase transition, $T\lambda^{k}_{m}(0)-1$, and that all coefficients grow from zero with this distance, except for $D^{*}$ which varies discontinuously through $T_{c}$. Furthermore, the normalized diffusion coefficient increases from zero with $\left<k_{-}k_{0}\right>\left<k_{0}k_{+}\right>$ and $\left<k_{-}k_{+}\right>$ -- the correlation moments of the degree distribution which encode the propensity for transport from the $i\mp1$ site to the $i\pm1$ site (both of which cannot be zero, otherwise epidemics are locally confined), and with the viability of subdominant modes to support growth. In general, we find that as  $\delta$ increases: $v^{*}$ and $D^{*}$ increase, $q^{*}$ decreases, and $s^{*}$ remains constant, implying faster transport and greater similarity among sites, as more edge-type pairs allow for traversing the lattice, but with little change in the growth exponent.

The above demonstrates the typical trend for these models, that the front dynamics is strongly influenced by the joint degree distribution's second moments (i.e., the relevant excess degree properties  are generally amplified by correlation among degree types and degree heterogeneity). For example, in analogy with the single network case, fast transport can be achieved with the presence of a few nodes with large internal and external degrees, or ``transport hubs", even if the average degrees in the network are small \cite{Newman3,Vespignani1}. 

\subsection{\label{sec:SimpleMixing} Simple Mixing Example} 
Additional understanding of the basic form of the transport coefficients is gained by looking at a special case of the micro-scale degree distribution,  where the generating function takes the form $G\left(fx_{0}+\frac{1-f}{2}(x_{+}+ x_{-})\right)$, with total degree described by $G$, and a given edge connecting nodes of the same site with probability $f$, and nodes of left and right neighboring sites with equal probability  $\left(1-f\right)/2$, where $1-f$ is an effective mixing parameter among populations. With this prescription, the critical transmissibility is reduced to the inverse of the total-edge excess degree, $T_{c}=\frac{G'(1)}{G''(1)}$, and the normalized diffusion coefficient, to the fraction of external edges in a each direction, $\delta=(1-f)/2$. Moreover, the dispersion relation takes the instructive form 
\begin{equation}
\label{eq:SimpleDisp}
s(q)= -1 + \frac{TG''(1)}{G'(1)}\left(f + (1-f)\cosh(q) \right) \;, 
\end{equation}
\noindent where $s+1$ is given by the basic reproductive number multiplied by the average relative incidence, $e^{-\Delta x  q}$, on the end of a randomly selected edge -- illustrating the intuitive generalization of the single network case, where different edge types are more and less likely to connect to infected nodes depending on their place in the lattice, and thus to contribute to local growth. 

\begin{figure}[t] 
\centerline{\includegraphics[scale=0.43]{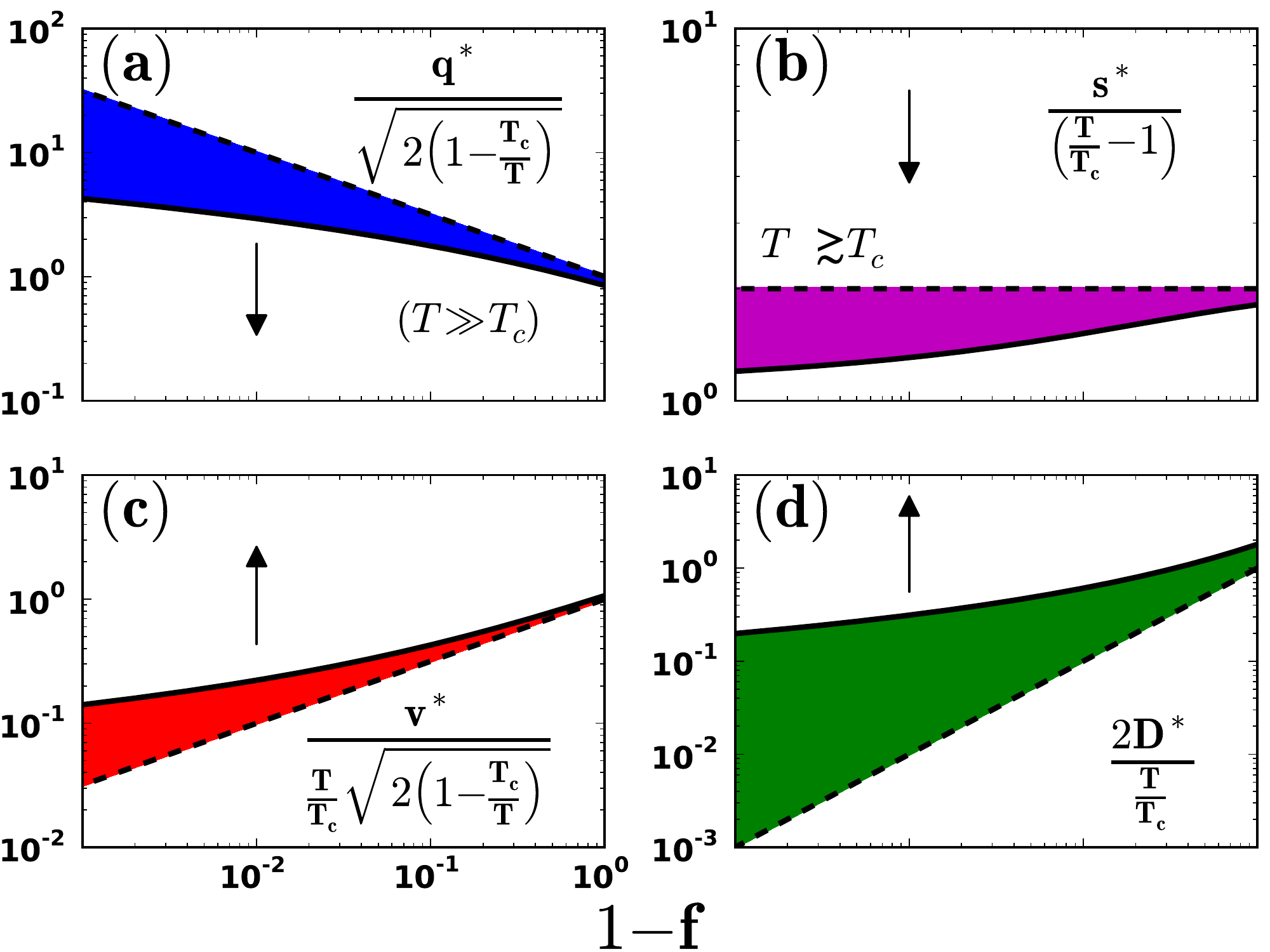}}
\caption{{{The scaled transport coefficients for a one-dimensional lattice of configuration model networks with arbitrary total-degree distribution and inter-population mixing parameter $1-f$, shown as functions of the latter (Sec.\ref{sec:SimpleMixing}): $q^{*}$ (a), $s^{*}$ (b), $v^{*}$ (c), and $D^{*}$ (d). The colored regions mark the range of each coefficient, which are bounded by the critical-region scaling $(T \gtrsim T_{c})$, and the limiting behavior $(T \gg T_{c})$,  delineated by dashed and solid curves respectively; the former are straight lines, signifying agreement with the predicted scaling \eqref{eq:Scaling}. Each panel's arrow indicates the direction of increase in the distance from the phase transition, $T/{T_{c}}-1$.
}}}
\label{fig:Transport}
\end{figure}

Likewise, from \eqref{eq:SaddlePoint} and \eqref{eq:Diff}, we find the speed and diffusion coefficient,  
\begin{equation}
\label{eq:SimpleSpeed}
v^{*}= \frac{TG''(1)}{G'(1)}(1-f)\sinh(q^{*}) 
\end{equation}
\noindent and 
\begin{equation}
\label{eq:SimpleDiff}
D^{*}= \frac{TG''(1)}{2G'(1)}(1-f)\cosh(q^{*}) 
\end{equation}
\noindent where $q^{*}$ satisfies \eqref{eq:SaddlePoint}, and $v^{*}$ is given by the basic reproductive number multiplied by the average product of relative position and incidence, $-\Delta x e^{-\Delta x  q^{*}}$, on the end of a randomly selected edge. Fig. \ref{fig:Transport} shows the transport coefficients, \eqref{eq:SimpleDisp}, \eqref{eq:SimpleSpeed}, and \eqref{eq:SimpleDiff}, as functions of $T/T_{c}$ and $f$, with partial scaling collapse \eqref{eq:Scaling} for the corresponding class of network configurations. The expected reaction-diffusion scaling can be observed near the critical point, and far away from the critical region, when $T \gg T_{c}$, $q^{*}$, $\frac{s^{*}G'(1)}{TG''(1)}$, $\frac{v^{*}G'(1)}{TG''(1)}$, and $\frac{D^{*}G'(1)}{TG''(1)}$ tend to limiting curves which depend only on $1-f$ , suggesting the intuitive asymptotic proportionality to the reproductive number. 

\subsection{\label{sec:PulledFront} Pulled Front Classification} 
In order to connect the transport properties of the linear equations to the full nonlinear system, we refer to the classification of fronts propagating into unstable states, which in our system is the fully susceptible metapopulation lying ahead of the infection front.  In general, there are two types of deterministic fronts: pulled and pushed, with the former having fronts with asymptotic speed equal to the linear spreading speed and the latter having fronts with asymptotic speed greater than the linear spreading speed \cite{Saarloos1}. Pushed fronts occur because nonlinearities in the equations of motion tend to increase the growth of perturbations on the unstable state, resulting in nontrivial front shape dependence of the speed. However, in our system all nonlinearities are proportional to probability generating functions, $\sim G'(\theta)/G'(1)$ \eqref{eq:1DLatt}, which are monotonically increasing over the unit interval. Therefore, all nonlinearities tend to increase $\theta$, and consequently dampen the growth of infection -- a sufficient condition for pulled fronts \cite{Panja}, and thus we anticipate fronts in this model to be pulled; this agrees with the intuition that epidemic propagation is governed by its behavior in a fully susceptible population. In practice, the classification has importance for control strategies in systems with similar structure, implying that to mitigate the spread of infection among populations, efforts should be focused on the leading edge of the front, and not on larger outbreaks occurring farther behind.

\subsection{\label{sec:LocalDynamics} Relaxation Properties}
In addition to quantifying the transport, the front speed can be used to extract information about the dynamics away from the unstable state. As shown above, the $\vec \theta$-field settles onto a solution with translational similarity, $\vec{\theta}_{n\pm x}(t)=\vec{\theta}_{n}(t\mp \frac{x}{v^{*}})$, after an initial transient period. Behind the leading edge of the front, the behavior resembles a relaxation to the stable equilibrium \eqref{eq:SteadyState}, $\vec{\theta}_{n}(t)\approx \vec{\bar{\theta}}+\vec{\eta}(t-\frac{n}{v^{*}})\approx \vec{\bar{\theta}}+\vec{\eta}e^{-z^{*}(n-v^{*}t)}$, where the spatial rate, $|z^{*}|$, is the dominant eigenvalue of the nonlinear eigenvalue equation
\begin{align}
\label{eq:Relax}
& \det\left(\underline{\underline{G'_{e}}}(\vec{\bar{\theta}},z^{*})-\frac{1\!+\!v^{*}z^{*}}{T}\underline{\underline{I}}\right)=0, \;\;\;\; \text{with} \\
& \underline{\underline{G'_{e}}}(\vec{\bar{\theta}},z) = \left. 
\renewcommand{\arraystretch}{2.4}
\setlength\arraycolsep{7.0pt}
\begin{bmatrix*}[l]
\textstyle{\frac{G_{00}}{G_{0}(\vec1)}} &\frac{G_{0+}}{G_{0}(\vec1)}&\frac{G_{0-}}{G_{0}(\vec1)} \\
\frac{G_{-0}}{G_{-}(\vec1)}e^{-z}&\frac{G_{-+}}{G_{-}(\vec1)}e^{-z}&\frac{G_{--}}{G_{-}(\vec1)}e^{-z} \\
\frac{G_{+0}}{G_{+}(\vec1)}e^{z}&\frac{G_{++}}{G_{+}(\vec1)}e^{z}&\frac{G_{+-}}{G_{+}(\vec1)}e^{z} \\
\end{bmatrix*}
\right \vert_{\mathlarger{\vec{\bar{\theta}}}} \;. \nonumber
\end{align}
\noindent The latter is the analogue of $\underline{\underline{K_{e}}}(q)$ at the stable state, which does not depend on the first two moments of the degree distribution directly, but on the generating function's properties near the equilibrium \eqref{eq:SteadyState}. In general, the two characteristic spatial rates for this system are not equal, $|z^{*}| \neq q^{*}$, and when their difference is large, it often signifies a significant separation in the time scales of growth, $1/s^{*}$, and relaxation $1/v^{*}|z^{*}|$. The latter provides an estimate for the amount of time a site is infectious, with $1/|z^{*}|$ yielding a related estimate for the width of the propagating front (i.e., the typical spatial extent of an outbreak at a given time). In particular, when the front speed is very fast and the degree distribution's second moments are large with the first moments  $\mathcal{O}(1)$, we find that $|z^{*}|\ll q^{*}$, which suggests broad front profiles. In this case, propagation and relaxation can be thought of as approximately distinct processes.  
\begin{figure}[th]
\includegraphics[scale=0.47]{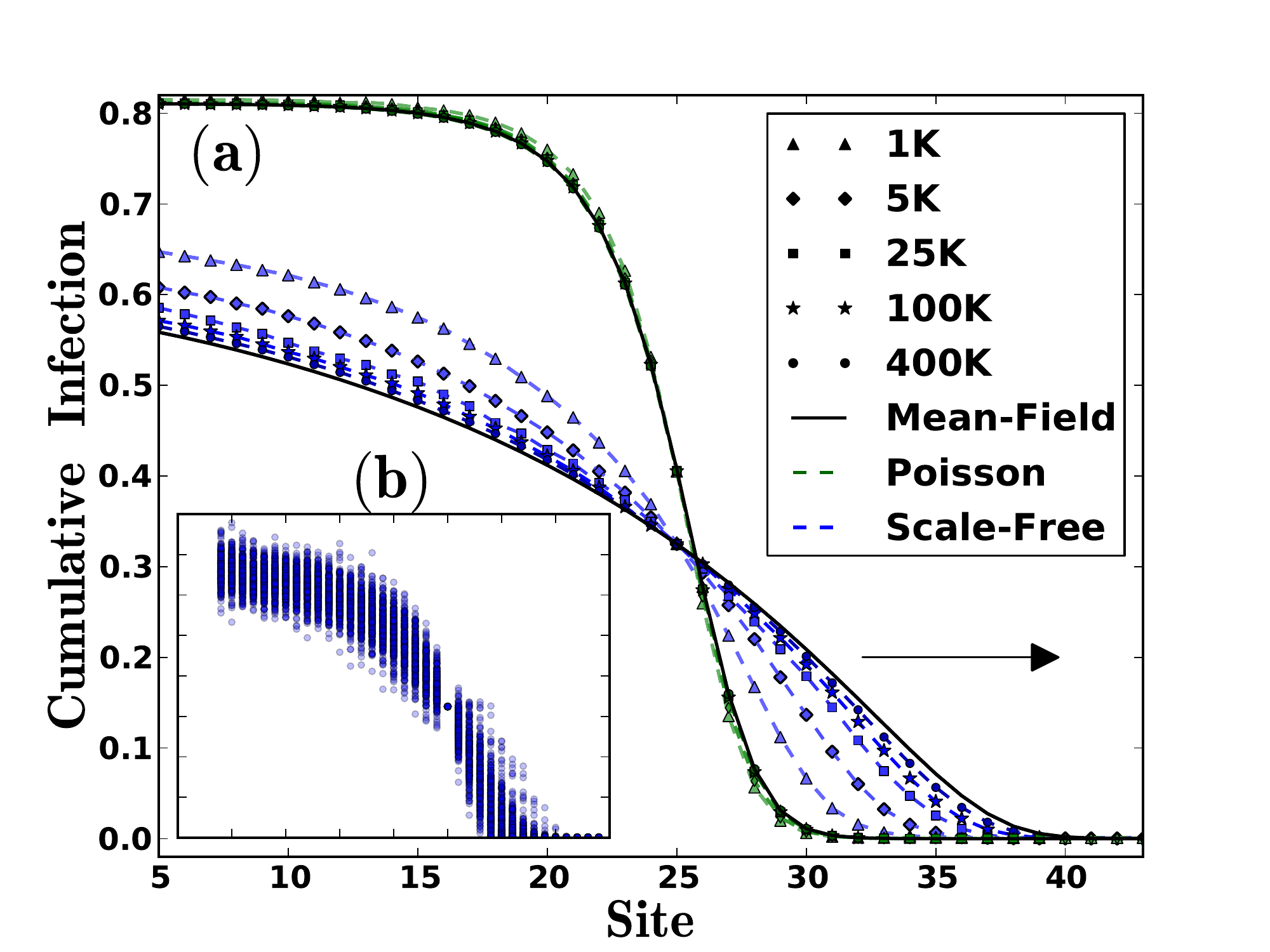}
\caption{{{(a) A comparison between the average cumulative infection profile for stochastic simulations of SIR dynamics on a one-dimensional lattice of scale-free (blue) and Poisson (green) networks and mean-field predictions \eqref{eq:1DLatt}. Various site sizes are shown with different symbols and color shades -- varying from light to dark for $10^{3}$ to $4\times10^{5}$ respectively. Front shapes for increasingly large sizes are found to converge to the respective mean-field front. The parameters for the underlying graphs were chosen to be: $K=100$, $\alpha=2$ and $p=0.3$ for the scale-free, and $C=2.90157$ and $p=0.3$ for the Poisson (Sec.\ref{sec:Compare}). A lattice size of 50 sites was used, which was large enough to ensure uniformity with the above graph parameters and reaction rates $\beta=\gamma=1$. The arrow indicates the propagation direction. (b) Stochastic front realizations conditioned on the middle lattice site having cumulative infection equal to half the equilibrium value \eqref{eq:SteadyState}. Averaging over realizations produced profiles like those in (a).}}} 
\label{fig:Profile}
\end{figure} 
\section{\label{sec:Compare} COMPARISON WITH STOCHASTIC SIMULATIONS}

The above predictions for the mean-field dynamics on the one-dimensional metapopulation were compared to stochastic simulations of SIR dynamics on random instances of multi-scale, metapopulation networks, using Gillespie's Direct Method \cite{Keeling1, Gillespie1, Gibson}. The graphs were constructed using the multitype configuration model by first generating a degree sequence from the desired degree distribution and then connecting pairs of edge-``stubs", selected uniformly at random \cite{Newman3, Antoine1,Newman1}.  An outbreak was started by choosing one node from the first lattice site to be infected with all others susceptible. Only outbreaks which lead to epidemics with $\mathcal{O}(N)$ cumulative infection were considered for comparison with mean-field predictions. In order to ignore fluctuations in the initial transients, time was zeroed after the first 100 reactions.

\begin{figure}[t]
\centerline{\includegraphics[scale=0.46]{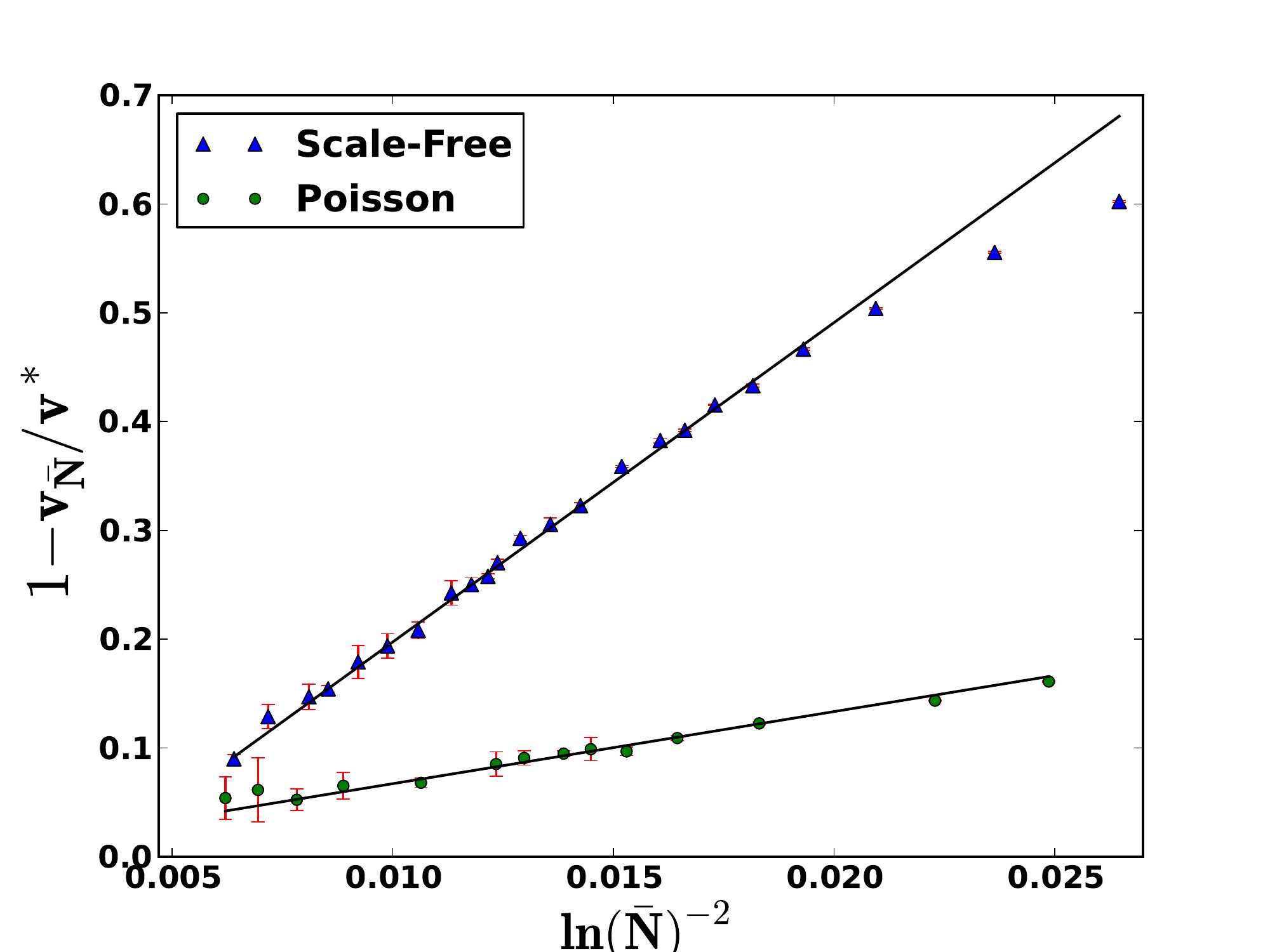}}
\caption{{{(a) Convergence of the average velocities, $v_{\bar{N}}$, to the mean-field predictions, $v^{*}$, for scale-free (blue) and Poisson (green) networks (Fig. \ref{fig:Profile}) as functions of the cumulative number of infected nodes at each site, shown with fits to the expected pulled front scaling,  $v^{*}-v_{\bar{N}} \sim \frac{D^{*}q^{*}\pi^{2}}{\ln^{2}(\bar{N})}$(Sec.\ref{sec:Compare}).}}} 
\label{fig:VelocityConvergence}
\end{figure} 

We are interested in the average shape of the front that connects the fully susceptible unstable state lying ahead of the infectious wave and the fully recovered (equilibrium) state lying behind it. 
The average shape was computed by taking instantaneous ``snapshots'' of the profile for each stochastic realization, conditioned on the middle lattice site having cumulative infection equal to half the equilibrium value  \eqref{eq:SteadyState}, and averaging the cumulative infection of the other sites over different realizations. In general, the ``snapshots'' did not occur at the same instant; however, the shifting of different front realizations in time, such that they overlapped at a given point, and conditionally averaging over the shape, eliminated some of the effects of diffusive wandering. The measured fronts were compared to the mean-field  profiles by integrating the lattice equations \eqref{eq:1DLatt}. A comparison is shown in Fig. \ref{fig:Profile} for two graphs with scale-free and Poisson degree distributions, with generating functions
\begin{equation}
G_{S.F.}(\vec{x})= \textrm{Li}_{\alpha}\left(e^{-1/K}x_{0}\left(1-p+px_{+}\right)\left(1-p+px_{-}\right)\right) \nonumber \\
\end{equation}
\noindent and
\begin{equation}
G_{P}(\vec{x})=  \exp \left(C\left(x_{0}\left(1-p+px_{+}\right)\left(1-p+px_{-}\right)-1\right) \right) \;, \nonumber 
\end{equation}
\noindent where $\textrm{Li}_{\alpha}$ is the polylogarithm function with exponent $\alpha$ \cite{Newman3}. The parameters for the degree distributions were chosen such that each network had the same average degree and cloning parameter, $p$ (i.e.,  given a specified internal degree distribution, each of a node's internal edges is copied to form an external edge with probability $p$), but with different inherent levels of heterogeneity. 

We see in Fig. \ref{fig:Profile} that the epidemic front is broader in the scale-free network than in the Poisson. This difference comes from the much larger front speed of the former, which had average excess degrees an order of magnitude larger than the latter, \eqref{eq:Dispersion} and  \eqref{eq:SaddlePoint}, and the relatively similar relaxation times \eqref{eq:Relax} for the two classes of networks (implying that the time scale over which a site is infectious in each network is roughly the same). In the more homogeneous Poisson networks, the front is more narrow and propagates through the lattice on the same time scales as the local infection dynamics; whereas in the scale-free case, the leading edge of the front propagates quickly through the lattice, followed by a slower relaxation to the stable equilibrium state behind the front. This comparison shows that assumptions of homogeneity can drastically underestimate the speed and extent of fronts in systems with heterogenous interactions. 

Additionally, the front speed $v$ was numerically estimated from the average time $\left<\tau_{prog}\right>$ required for the leading edge of the front to move forward by one lattice site (where the leading edge was defined as that site where the incidence first reached a set $\mathcal{O}(1)$ level) and averaging over such levels; i.e., $1/v=\left<\tau_{prog}\right>$, once the initial spatial variation had decayed. Fig. \ref{fig:VelocityConvergence} shows the convergence of the measured speed from simulations to the mean-field prediction for each graph as a function of the steady state, cumulative infected population size at every lattice site,  $\bar{N}=\bar{P}N$ (Sec.\ref{sec:VolzMiller}), with total size $N$. The lines represent fits to the expected scaling of the largest finite-size correction for pulled fronts, $v^{*}-v_{\bar{N}} \sim \frac{D^{*}q^{*}\pi^{2}}{\ln^{2}(\bar{N})}$, obtained from a general $1/\bar{N}$ cutoff in the mean-field equations  \cite{Panja, Saarloos1, Brunet}; the coefficients are found to be $\mathcal{O}(1)$ of  the expected scaling. In general, higher order corrections in $\bar{N}$ must be calculated  from an analysis of the full, stochastic system \cite{Panja}. The very slow convergence in $\bar{N}$ comes from the transport dependence on the linearized equations where infinitesimal infection levels apply and sensitivity to stochastic effects is highest. This can be seen in the fairly large finite-size corrections to the velocity, particularly for the scale-free network, leading to a more narrow conditionally averaged front relative to the mean-field, with fewer sites initiated at a given time (Fig. \ref{fig:Profile}). 

\begin{figure}[t]
\centerline{\includegraphics[scale=0.46]{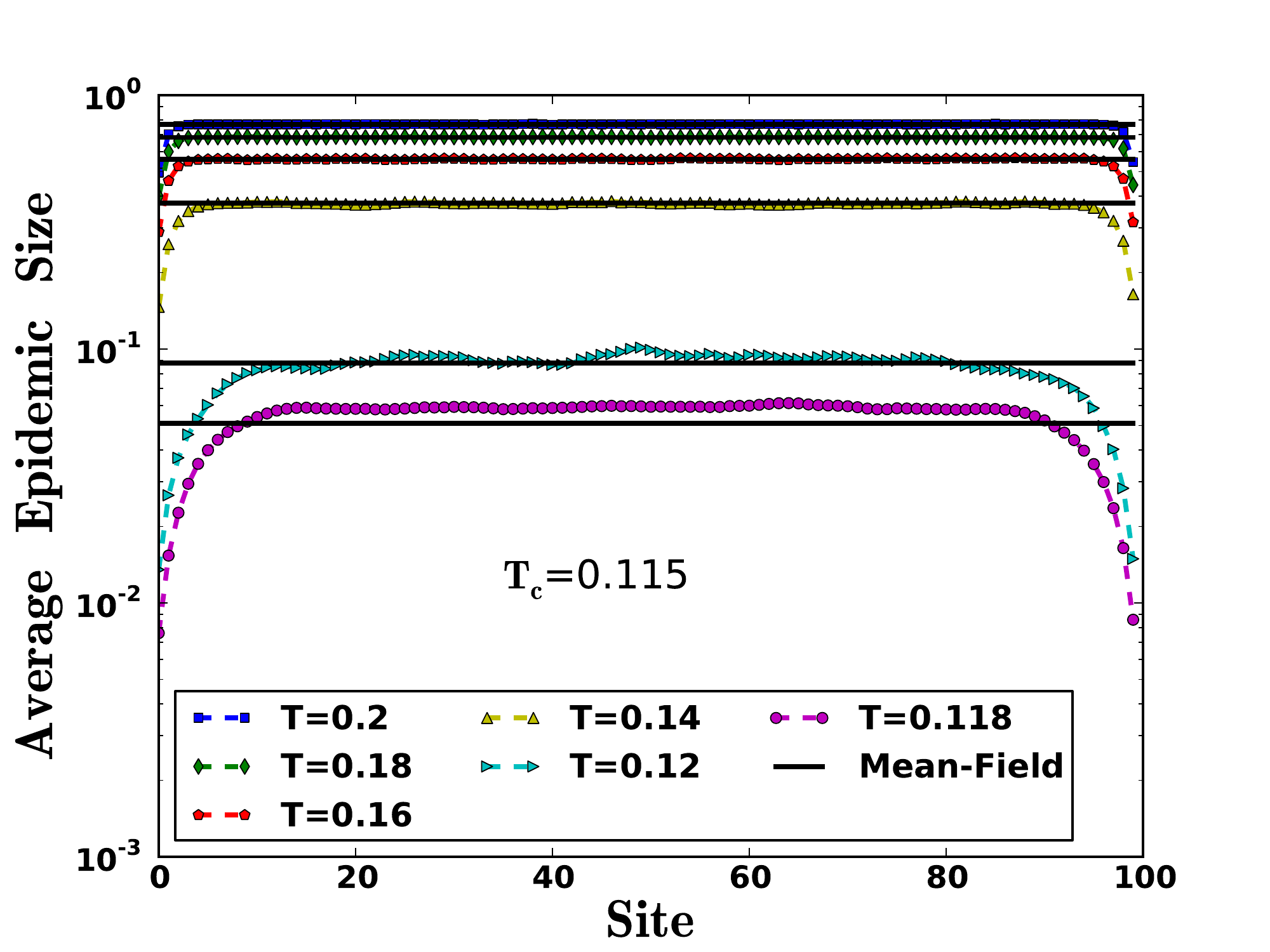}}
\caption{{{The average epidemic fraction at each site for a network with asymmetric  generating function (Sec.\ref{sec:Compare}) and varying transmissibilities. The critical point at which the epidemic vanished agrees with the prediction, $T_{c}=0.115$ \eqref{eq:1DThresh}.  Each site was occupied by 20,000 nodes on a lattice of 100 sites.}}}
\label{fig:Threshold}
\end{figure}  

Finally, the average epidemic profile and transmissibility threshold, \eqref{eq:SteadyState} and \eqref{eq:1DThresh}, were compared to simulations. Fig. \ref{fig:Threshold} plots those comparisons for a system with left-right asymmetric generating function 
\begin{equation}
G_{Asym}(\vec{x})=\frac{1}{3}\left( 2x_{0}^{2}x_{+}^{2}x_{-}^{4} +x_{0}^{4}x_{+}^{6}x_{-}^{2}\right) \;, \nonumber
\end{equation}
\noindent on a lattice of 100 sites, with 20,000 nodes on each site. Both the epidemic size for various transmissibilities and the threshold were found to be in good agreement with mean-field predictions, though the finite-size effects became increasingly important as the critical region was approached, leading to significantly smaller outbreaks near the edges of the lattice.
\section{\label{sec:Conclusion} CONCLUSION}
In this paper we have generalized a mean-field theory for infection dynamics on multitype networks, and used such networks to model multiscale metapopulations. Together, this enabled us to explore how macro-scale disease propagation depends on micro-scale interaction structure. As a necessary first step in this direction, we applied the approach to a simple metapopulation model for a chain of coupled populations, and derived the transport properties for infection, including their scaling with the disease transmissibility and the statistical properties of the underlying network. We also found a threshold for the viability of epidemics, and calculated the relaxation properties of the propagating front. These were compared for different network models, with heterogeneous networks having considerably higher speeds and broader fronts than their homogeneous counterparts -- illustrating the importance of including complexity in the fine-scale topology in order to accurately capture transport phenomenology. 

Various extensions of the work presented -- both in terms of analyses carried out and systems studied -- could be considered. We have addressed here only the average dynamics of the one-dimensional, homogeneous system, without any description of finite-size fluctuations, or consideration of the dynamics in higher-dimensional generalizations. Greater complexity could be introduced through the spatio-temporal dependence of network parameters, and/or more general network configurations \cite{Karrer}. An interesting extension of the model discussed here would include dynamic contacts between nodes and explicit mobility, instead of the assumed time scale separation between topology and the overlying process \cite{Perra,Belik1, Barabasi}. However, the basic formalism presented here can enable one to study such factors and build more realistic models for infectious processes in multiscale problems. 

\section*{\label{sec:Ack}ACKNOWLEDGMENTS}
This work was supported by the Science and Technology Directorate of the U.S.  Department of Homeland Security via the interagency agreement no.\ HSHQDC-10-X-00138. We thank Drew Dolgert for his assistance with our implementation of stochastic simulations.

\end{document}